\pdfoutput=1

\documentclass[final,3p,times,twocolumn]{elsarticle}

\usepackage{amssymb}
\usepackage{amsthm}
\usepackage{amsfonts}

\usepackage{dcolumn}
\usepackage{booktabs}
\usepackage{bm}

\newcolumntype{d}[1]{D{.}{.}{#1} }

\journal{Physics Letters B}

\begin{document}
\begin{frontmatter}



\title{A region of high-spin toroidal isomers}


\author[umcs]{Andrzej Staszczak\corref{cor1}}\ead{stas@tytan.umcs.lublin.pl}

\author[ornl]{Cheuk-Yin Wong}\ead{wongc@ornl.gov}

\cortext[cor1]{Corresponding author}
\address[umcs]{Institute of Physics, Maria Curie-Sk{\l}odowska University,
pl. M. Curie-Sk{\l}odowskiej 1, 20-031 Lublin, Poland}
\address[ornl]{Physics Division, Oak Ridge National Laboratory,
P.O. Box 2008, Oak Ridge, Tennessee 37831, USA}

\begin{abstract}
The combined considerations of both  the bulk liquid-drop-type behavior
and the quantized angular momentum reveal that high-spin toroidal isomeric states
may have general occurrences for light nuclei with 28$\leqslant$$A$$\leqslant$48.
High-spin  $N$=$Z$  toroidal isomers  in this mass region
have been  located theoretically using cranked self-consistent constraint
Skyrme-Hartree-Fock model calculations.
\end{abstract}


\end{frontmatter}





Nuclei as we now know them have sphere-like geometry.
Wheeler suggested that under appropriate conditions
the nuclear fluid may assume a toroidal shape
\cite{Gam61,Won72PL,Won72}.
Toroidal nuclei are however plagued with various instabilities
\cite{Won72}, and the search remains elusive \cite{Roy96,Zhe07,Sta08}.
It was found previously from the liquid-drop model that a ``rotation''
about the symmetry axis with an angular momentum $I$=$I_z$ above
a threshold can stabilize the toroidal nucleus and can lead
to a high-spin isomer \cite{Won78}.

The toroidal high-spin isomer, whose large angular momentum
$I$=$I_z$ must be  generated by the alignment of individual
nucleon angular momenta along the symmetry axis  \cite{Boh81},
provides an elegant example in quantum mechanics as how
an axially-symmetric system can acquire a quantized
angular momentum.
Furthermore, the nuclear fluid in the toroidal isomeric state
may be so severely distorted by the change from sphere-like geometry
to the toroidal shape that it may acquire bulk properties of its own,
to make it a distinct type of quantum fluid.
Finally, the toroidal high-spin isomer may be a source of energy,
as its decay to the ground state can release a large amount
of excitation energy and the possibility of toroidal high-spin
isomers may stimulate also future reaction studies to explore
their production and detection by fusion
of two ions at high angular momenta \cite{Eud12,Esb12}.
For all these reasons, the investigation on toroidal high-spin
isomers is of general interest.

In the liquid-drop model of a toroidal nucleus,
we can select the major radius $R$, the minor radius $d$,
the angular momentum $I$=$I_z$ about the symmetry axis,
and the corresponding rigid-body moment of inertia  $\Im_{\rm rigid}$
as macroscopic variables.
(For a sketch of $R$ and $d$, see Fig. 1 of \cite{Won72}.)
The energy $I(I+1)/2\Im_{\rm rigid}$ associated with the
angular momentum $I$ can be called the ``rotational'' energy.
The variation of the rotational energy and the Coulomb energy  tend to
counter-balance the variation of the  surface energy  \cite{Won78}.
As a consequence, there is an $I$-threshold above which the rotating toroidal nucleus can
be stable against a variation of $R/d$. The toroidal nucleus is
stable also against axially-asymmetric sausage distortions within an
$I$-window, when the same mass flow is maintained across the toroidal
meridian.  Beyond the $I$-window, sausage instabilities of higher
orders dominate to break the toroid into many beads \cite{Won78}.

To study toroidal high-spin states theoretically, we need a systematic
way to determine the quantized $I$ value, which is a non-trivial
function  of $N$ and $Z$.
The quantized $I$ can be obtained from the single-particle state
diagrams under the constraint of a fixed aligned angular momentum.
For simplicity, we limit our present studies to
even-even $N$=$Z$ nuclei.  Previously, an investigation of $^{40}$Ca
as the evolution of a chain of 10 alpha particles revealed that
$^{40}$Ca with $I$=60 $\hbar$ may represents a toroidal high K-isomeric state
\cite{Ich12}, in qualitative agreement with the $I$-threshold and $I$-window
concepts  in \cite{Won78}.

Accordingly, we need the energy diagram of the single-particle states
in a toroidal nucleus for different aligned angular momenta $I$.
For $I$=0 $\hbar$,  the single-particle potential for a nucleon in
a toroidal nucleus with azimuthal symmetry in cylindrical coordinates
$(r,z)$ can be represented by \cite{Won72}
\begin{equation}
V_0(r,z) = \frac{1}{2} m \omega_0^2 (r -R)^2 +\frac{1}{2} m \omega_0^2  z^2,
\end{equation}
where $\hbar\omega_0$=$[(3\pi R/2d)^{1/3}41/A^{1/3}]\langle\rho_{\rm
  torus}\rangle /\langle\rho_0 \rangle$.  We have included the ratio
$\langle\rho_{\rm torus}\rangle/\langle\rho_0 \rangle$ where
$\langle\rho_{\rm torus}\rangle$ and $\langle\rho_0\rangle$ are the
average nuclear densities in the toroidal and the spherical
configurations respectively, because the mean-field potential is
proportional approximately to the nuclear density.  In microscopic
calculations, $\langle \rho_{\rm torus}\rangle /\langle\rho_0\rangle$
is found to be approximately 1/2 to 2/3.  For $R\gg d$ and low-lying
states with the radial nodal quantum number $n_{\rho}$=0 and the
azimuthal nodal quantum number $n_z$=0, the expectation value of the
spin-orbit interaction is approximately zero \cite{Won72}, and we can
neglect the spin-orbit interaction.

We label a state by
$(n\Lambda \Omega\Omega_z)$, where $n$=$(n_{z}+n_{\rho})$, $\pm
\Lambda$ is the $z$-component of the orbital angular momentum, and
$\Omega=|\Lambda \pm 1/2|$ is the single-particle total angular
momentum with $z$-components $ \Omega_z$=$\pm \Omega$.  For $R\gg d$,
the single-particle energy of the $(n_{\rho} n_z \Lambda \Omega)$
state with  $I$=0 $\hbar$ is therefore
\begin{equation}
E(n \Lambda \Omega)
\sim \hbar \omega_0 (n +1) + \frac{\hbar^2\Lambda^2}{2 m R^2} .
\end{equation}
Fig.\ 1(a) gives the single-particle state energies as a function
of $R/d$ for a toroidal nucleus with $I$=0 $\hbar$.

\begin{figure}[htb]
\begin{center}
\includegraphics[width=\columnwidth]{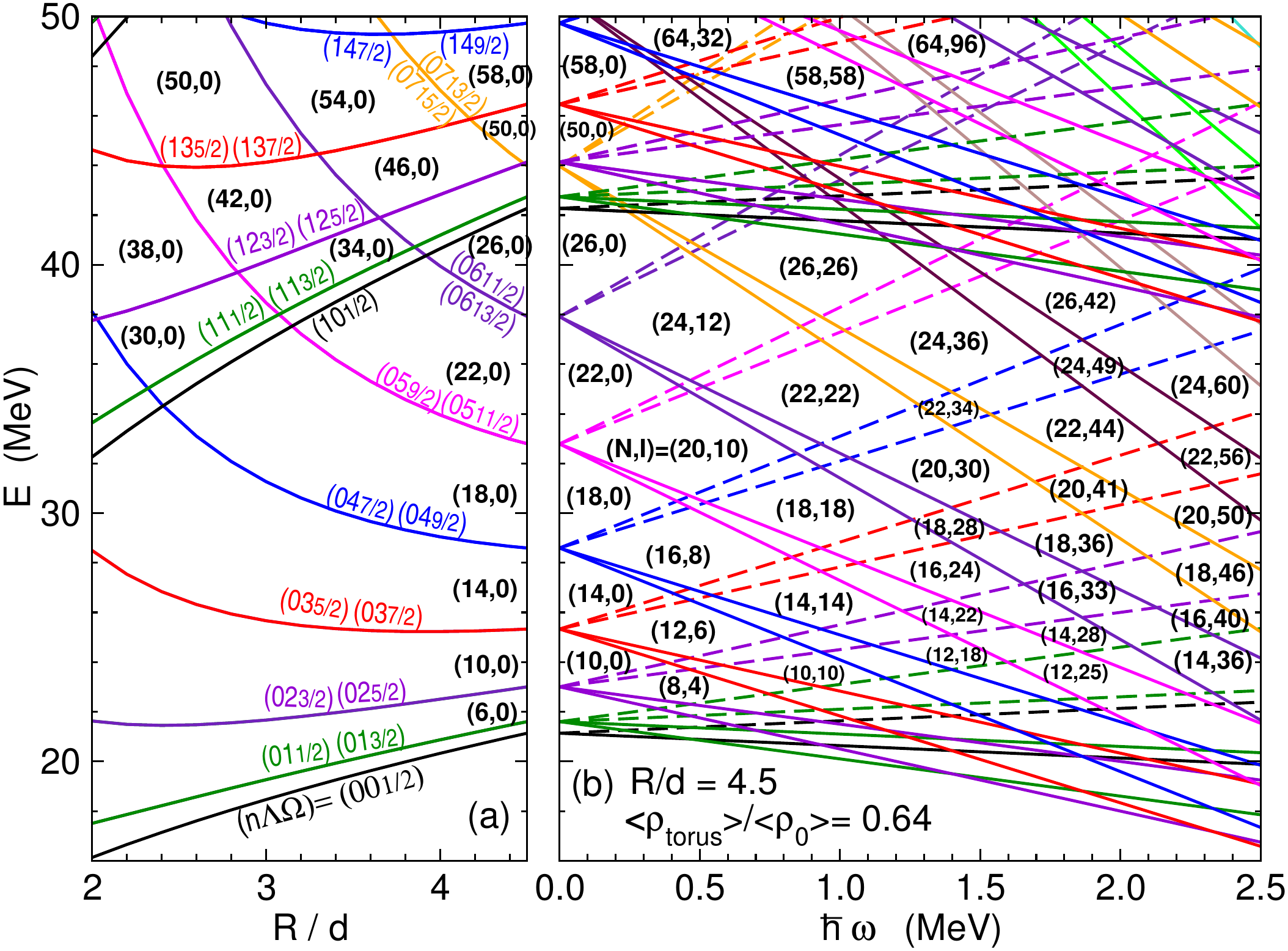}
\caption{\label{Fig1} (Colour online.)
  (a) Single-particle states of a toroidal nucleus with
  $I$=0 $\hbar$ as a function of  $R/d$, calculated with
  $\langle \rho_{\rm torus}\rangle /\langle \rho_0\rangle$=0.64
  and $A$$\sim$40. Each state is labeled by $(n, \Lambda, \Omega)$,
  with $n$=$n_z$+$n_\rho$ and degenerate $\Omega_z$.
  (b) Single-particle states of a toroidal nucleus with $R/d$=4.5,
  as a function of a cranking frequency $\hbar \omega$.
  Energy levels of positive- and negative-$\Omega_z$ states
  are given by the solid and dashed lines, respectively.
  The listed pair numbers $(N,I)$ refer to the
  occupation number $N$ and the total angular momentum $I$=$I_z$
  aligned along the symmetry $z$-axis.}
\end{center}
\end{figure}

For a non-collectively rotating toroidal nucleus with
aligned angular momentum,  $I$=$I_z$, we use
a Lagrange multiplier $\omega$ to describe the constraint
$I_{z}$=$\langle \hat{J}_{z} \rangle$=$\sum_{i=1}^{N} \Omega_{zi}$.
The constrained single-particle Hamiltonian becomes
$\hat{h}'={\hat h}-\omega \hat{J}_{zi}$, and the aligned angular
momentum $I$ is a step-wise function of the
Lagrange multiplier $\omega$ \cite{Ring80}, with each $I$
spanning a small region of $\hbar \omega$.  As the constrained
Hamiltonian $\hat{h}'$ is of the same form as that of a nucleus under
an external cranking, the constraint can be effectively described as a
cranking of the nucleus with an angular frequency $\omega$
\cite{deV83,Nil95}.  The single-particle state energy of the
$(n \Lambda \Omega \Omega_z)$ state, under the constraint
of the non-collective aligned angular momentum $I$ is
\begin{equation}
E(n \Lambda \Omega \Omega_z)
\sim \hbar \omega_0 (n +1) + \frac{\hbar^2\Lambda^2}{2 m R^2}
-\hbar \omega \Omega_z.
\end{equation}
Fig.~\ref{Fig1}(b) gives the single-particle state energies as a
function of the constraining Lagrange multiplier $\hbar \omega$,
for a toroidal nucleus with $R/d$=4.5, approximately the aspect
ratio for many toroidal nuclei in this region.  We can use
Fig.~\ref{Fig1}(b) to determine $I$=$I_z$ as a function of $N$ and
$\hbar \omega$.  Specifically, for a given $N$ and $\hbar \omega$, the
aligned $I_z$-component of the total angular momentum $I$ from the $N$
nucleons can be obtained by summing $\Omega_{zi}$ over all states
below the Fermi energy.

There are shell gaps for different $(N,I_{z})$ configurations in
Fig.~\ref{Fig1}.  They represent configurations with relative
stability for which additional shell corrections on top of the
liquid-drop-type energy surface \cite{Bra72,Won72} may enhance the
stability for toroidal configurations.  The energy scales of the
$\hbar \omega$ and $E$ axes in Fig.~\ref{Fig1}(b) depend on $N$,
$R/d$, $\langle \rho_{\rm torus} \rangle/ \langle \rho_0 \rangle$
which vary individually at different isomeric toroidal energy minima,
but the structure of the $(N,I_{z})$ shells and their relative
positions in Fig.~\ref{Fig1}(b) remain approximately the same in this
$A$$\sim$40 mass region.  We can use Fig.~\ref{Fig1}(b) as a
qualitative guide to explore the landscape of the energy surface for
different $(N,I_{z})$ configurations, by employing a reliable
microscopic model.

A microscopic theory that includes both the single-particle shell
effects and the bulk properties of a nucleus is the Skyrme energy
density functional approach in which we solve an equality-constrained problem:
\begin{equation}
\left\{
\begin{array}{l}
\displaystyle\min_{\boldsymbol{\bar{\rho}}} E^{tot}[\boldsymbol{\bar{\rho}}]\\
         \mbox{subject to: } \displaystyle\langle \hat{N}_{q} \rangle= N_{q},\\
\phantom{\mbox{subject to: }}\displaystyle\langle \hat{Q}_{\lambda\mu} \rangle =Q_{\lambda\mu},\\
\phantom{\mbox{subject to: }}\displaystyle\langle \hat{J}_{i} \rangle =I_{i},
\end{array}
\right. 
\label{eq:3}
\end{equation}
where an objective function,
$E^{tot}[\boldsymbol{\bar{\rho}}]=\langle \hat{H}_{Sk} \rangle$,
is the Skyrme energy density functional  \cite{Vau72}.
The constraint functions are defined by average values of the
proton/neutron particle-number operator, $\hat{N}_{p/n}$, the
mass-multiple-moment operators, $\hat{Q}_{\lambda\mu}$, and the
components of the angular momentum operator $\hat{J}_{i}$.
$N_{p/n}=Z/N$ are the proton/neutron numbers, $Q_{\lambda\mu}$ are the
constraint values of the multiple-moments, and $I_{i}$ are the
constraint components of the angular momentum vector.

\begin{figure}[htb]
\begin{center}
  \includegraphics[width=\columnwidth]{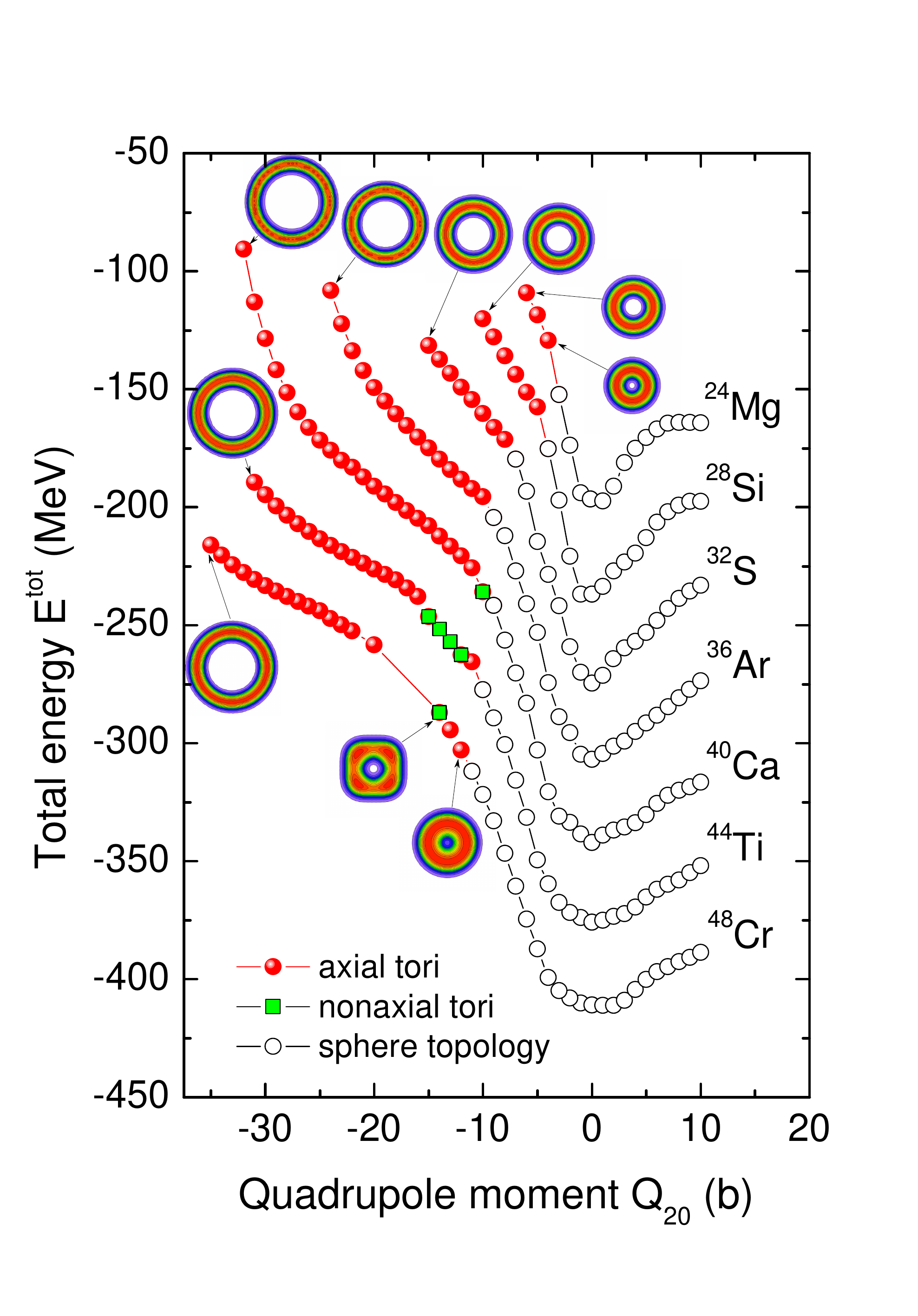}
  \caption{\label{Fig2} (Colour online.) The total HFB energy of
    $^{24}$Mg, $^{28}$Si, $^{32}$S, $^{36}$Ar, $^{40}$Ca, $^{44}$Ti,
    and $^{48}$Cr as a function of the quadrupole moment for the case of $I$=0.
    Axially-symmetric toroidal configurations are indicated
    by the solid bullet points, axially-asymmetric toroidal
    configurations by squared points, and configurations with the
    topology of a sphere by open circles.  Some toroidal density
    distributions are displayed.}
\end{center}
\end{figure}

The above  constraint equations were solved using an augmented
Lagrangian method \cite{alm} with the
symmetry-unrestricted code HFODD \cite{hfodd}.  In the particle-hole channel the Skyrme
SkM* force \cite{Bar82} was applied and a density-dependent mixed
pairing \cite{Dob02,Sta09} interaction in the particle-particle
channel was used. The code HFODD uses the basis expansion method in a
three-dimensional Cartesian deformed harmonic oscillator basis.  In
the present study, we used the basis which consists of states having
not more than $N_{0}$=26 quanta in the Cartesian directions, and not
more than 1140 states.

Our objective is to locate local toroidal figures of equilibrium, if
any, in the multi-dimensional search space of $(A,Q_{20},I)$.  We
first map out the energy landscape for axially-symmetric toroidal
shapes under these $Q_{20}$ and $I$ constraints, with fine grids in
$Q_{20}$ and all allowed non-collective rotations in 0 $\leqslant
I\leqslant$ 120 $\hbar$ for different $A$.  If the topographical landscape
reveals a local energy minimum then the quadrupole constraint is
removed at that minimum and free-convergence is tested to
ensure that the non-collectively rotating toroid nucleus is indeed
a figure of equilibrium.

For the case of $I$=0, as shown in Fig.~\ref{Fig2}, the
Skyrme-Hartree-Fock-Bogoliubov (HFB) calculations for $N$=$Z$ with
24$\leqslant$A$\leqslant$48 reveal that as the quadrupole moment
constraint, $Q_{20}$, decreases to become more negative, the density
configurations with sphere-like geometry (open circular points)
turn into those of an axially-symmetric torus (solid bullet points),
as would be expected from the single-particle state diagrams of
Fig.~\ref{Fig1}(a). The energies of axially-symmetric toroidal
configurations as a function of $Q_{20}$ lie on a slope.  This
indicates that even though the shell effects cause the density to
become toroidal when there is a quadrupole constraint, the magnitudes
of the shell corrections are not sufficient to stabilize the tori
against the bulk tendency to return to sphere-like geometry.

\begin{figure}[htb]
\begin{center}
  \includegraphics[width=\columnwidth]{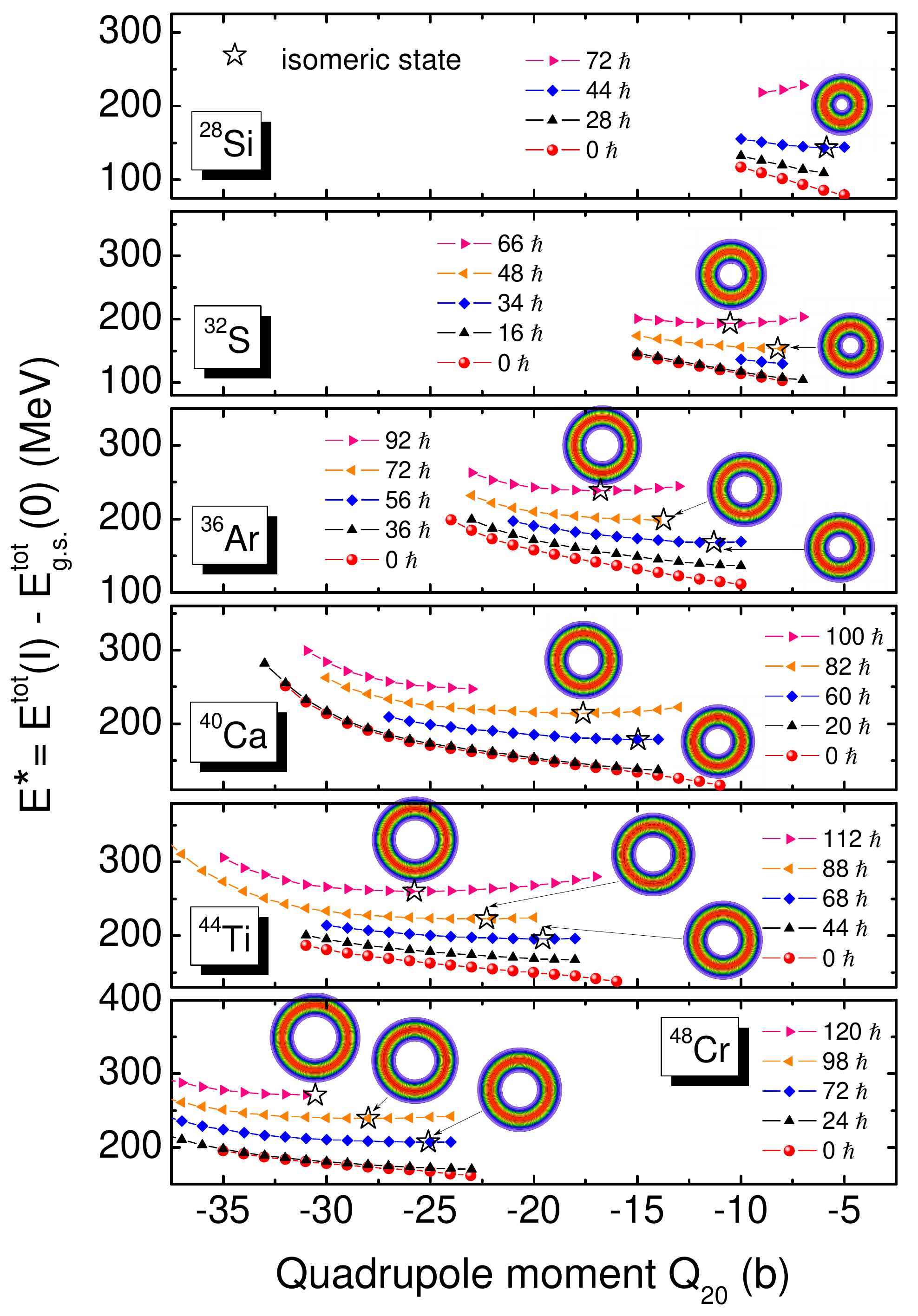}
  \caption{\label{Fig3} (Colour online.) The excitation energy of high-spin
    toroidal states ($E^*$) of
    $^{28}$Si, $^{32}$S, $^{36}$Ar, $^{40}$Ca, $^{44}$Ti, and
    $^{48}$Cr as a function of $Q_{20}$ for different angular momentum
    along the symmetry axis, $I$=$I_z$. The density
    distributions and locations of isomeric toroidal energy minima are
    indicated by the star symbols.}
\end{center}
\end{figure}

We next extend our Skyrme-HFB calculations further to include both the
quadrupole moment $Q_{20}$ constraint and the angular momentum
constraint, $I$=$I_z$. The pairing energies are smaller for
toroidal nuclei than with a spherical geometry, for a case of $I$=0,
additionally pairing interaction is suppressed as the two degenerate
$\pm \Omega_z$ states split apart under the constraining  $\hbar \omega$
when $I$$\neq$0.
We shall carry out the cranking calculations without
the pairing interaction, using a Skyrme-HF approach.
The results of such calculations for 28$\leqslant$$A$$\leqslant$48
are presented in Fig.~\ref{Fig3}, where we plot the
excitation energy of the high-spin toroidal states relative to the spherical
ground state energy, $E^{*}=E^{\rm tot}(I)-E^{\rm tot}_{\rm g.s.}(0)$,
as a function of the constrained $Q_{20}$, for different quantized
$I$.  For each point $( Q_{20},I)$ on an $I$ curve for a fixed $A$, it
was necessary to adjust $\hbar \omega$ judiciously within a range to
ensure that the total aligned angular momentum of all nucleons in the
occupied states gives the quantized $I$ value of interest. The energy
curves in Fig.~\ref{Fig3} become flatter as $I$ increases, similar to
the energy curves in the liquid-drop model as the angular momentum
increases \cite{Won78}.

With our systematic method outlined above, we are able to locate many high-spin
toroidal isomeric states:
$^{28}$Si($I$=44 $\hbar$),
$^{32}$S($I$=48, 66 $\hbar$), $^{36}$Ar($I$=56, 72, 92 $\hbar$), $^{40}$Ca($I$=60,
82 $\hbar$), $^{44}$Ti($I$=68, 88, 112 $\hbar$), and $^{48}$Cr($I$=72, 98, 120 $\hbar$),
as shown in Fig.~\ref{Fig3} and listed in Table I.
Note that with a fixed initial shape of a ring of 10 alpha particles,
the earlier result of \cite{Ich12} finds only a single case of
$^{40}$Ca($I$=60 $\hbar$) as an isomeric toroidal figure
of equilibrium.  However, with the help of Fig.~\ref{Fig1}(b) and the
fine grids in the large multi-dimensional space of ($A,Q_{20},I$), we
find a large number of isomers, demonstrating  the general occurrence
of toroidal high-spin states.
The $A$ and $I$ values have their correspondences in the $(N,I)$
shells in Fig.~\ref{Fig1}(b). The equilibrium configurations at the
energy minima have been tested and found to be self-consistently
free-converging after the removal of the quadrupole moment
$Q_{20}$ constraint.

Table~\ref{table1} gives the properties of the high-spin toroidal isomers in
28$\leqslant$$A$$\leqslant$48: their $Q_{20}$, $\hbar \omega$,
and excitation energy $E^*$ values, obtained with the Skyrme
SkM* interaction.  The excitation energy is of order 140-270 MeV.
The toroidal density can be parametrized as a Gaussian function,
$\rho(r,z)$=$\rho_{\rm max} \exp\{-[(r-R)^2+z^2]/(d^2 /\ln 2)\}$,
where $R$, $d$, and $\rho_{\rm max}$ for isomeric states are listed.
While the major radius $R$ and $R/d$ increase with increasing $A$,
the minor radius $d$ remains to be approximately the same.

\begin{table}[h]
\caption{\label{table1} Properties of high-spin toroidal isomers
  at their local energy minima in 28$\leqslant$$A$$\leqslant$48.
}
\resizebox{\columnwidth}{!} {
\begin{tabular}{crcd{3}ccccc}
\toprule
          & $I/\hbar$ & $Q_{20}$ & \multicolumn{1}{c}{$\hbar\omega$} &
$E^*$     & $R$       &   $d$    & $R/d$ & $\rho_{\rm max}$ \\
          &           & (b)      &  \multicolumn{1}{c}{(MeV)}        &
(MeV)     &~(fm)~     &~(fm)~&       & (fm$^{-3}$)  \\ \midrule
$^{28}$Si &        44 &  -5.86   &  2.8          & 143.18 & 4.33  & 1.45 & 2.99  & 0.119 \\
$^{32}$S  &        48 &  -8.22   &  1.9          & 153.87 & 4.87  & 1.42 & 3.43  & 0.122 \\
          &        66 & -10.51   &  2.2          & 193.35 & 5.57  & 1.40 & 3.98  & 0.108 \\
$^{36}$Ar &        56 & -11.31   &  1.7          & 168.03 & 5.44  & 1.40 & 3.88  & 0.125 \\
          &        72 & -13.73   &  1.85         & 198.63 & 6.04  & 1.39 & 4.34  & 0.113 \\
          &        92 & -16.78   &  2.0          & 238.56 & 6.73  & 1.37 & 4.91  & 0.103 \\
$^{40}$Ca &        60 & -14.96   &  1.5          & 178.36 & 5.97  & 1.40 & 4.26  & 0.126 \\
          &        82 & -17.61   &  1.9          & 214.23 & 6.51  & 1.39 & 4.68  & 0.117 \\
$^{44}$Ti &        68 & -19.57   &  1.2          & 195.46 & 6.55  & 1.39 & 4.71  & 0.128 \\
          &        88 & -22.27   &  1.4          & 223.09 & 7.01  & 1.38 & 5.08  & 0.120 \\
          &       112 & -25.76   &  1.6          & 260.24 & 7.56  & 1.37 & 5.52  & 0.113 \\
$^{48}$Cr &        72 & -25.08   &  1.2          & 207.12 & 7.12  & 1.38 & 5.16  & 0.128 \\
          &        98 & -28.00   &  1.4          & 239.26 & 7.54  & 1.37 & 5.50  & 0.122 \\
          &       120 & -30.55   &  1.43         & 271.02 & 7.90  & 1.36 & 5.81  & 0.118 \\
\bottomrule
\end{tabular}
}
\end{table}

\begin{figure}[htb]
\begin{center}
  \includegraphics[width=0.75\columnwidth]{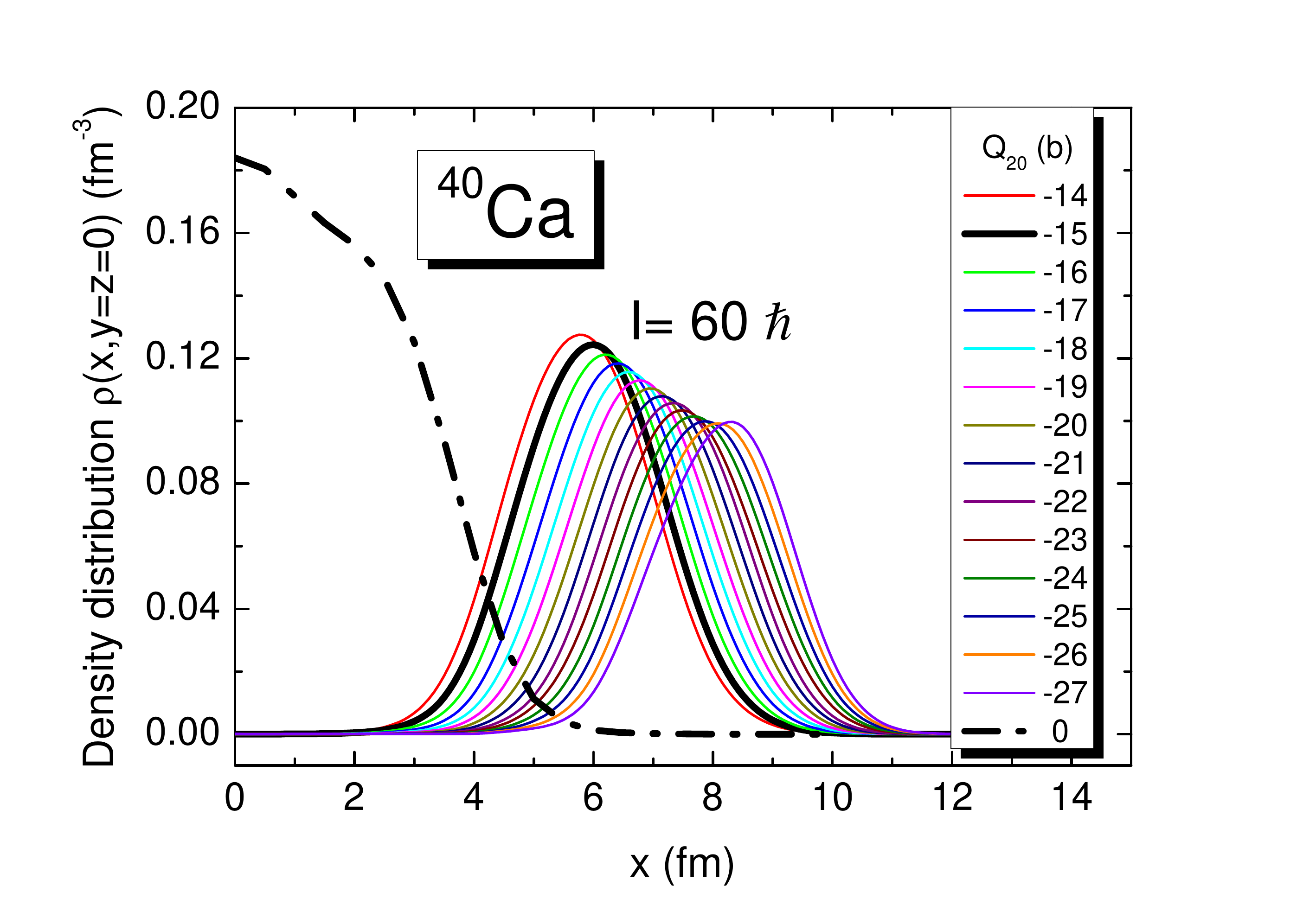}
  \caption{\label{Fig4} (Colour online.) The density distributions of
    $^{40}$Ca for $I$=60 $\hbar$ as a cut in the radial direction $x$ for
    different $Q_{20}$: (i) the thick solid curve is for the
    equilibrium toroidal configurations at the local energy minimum at
    $Q_{20}$=-15 b, (ii) the thin line curves are for the other
    $Q_{20}$ as labelled, and (iii) the dashed-dot curve is for
    $^{40}$Ca in the spherical ground state.}
\end{center}
\end{figure}

We plot in Fig.~\ref{Fig4} the density distributions of the toroidal
configurations of $^{40}$Ca with $I$=60 $\hbar$ as a cut in the radial
$x$-direction for different $Q_{20}$.  One notes that the average
density for $^{40}$Ca($I$=60 $\hbar$) at the toroidal energy minimum of
$Q_{20}=-15$ b (thick solid curve) is only 0.64 of the average nuclear
density for a spherical $^{40}$Ca (dash-dot curve).  This is a general
phenomenon for light toroidal nuclei, as the nuclear density is
affected by the presence of all forces \cite{Won85}.

\begin{figure}[htb]
\begin{center}
  \includegraphics[width=0.75\columnwidth]{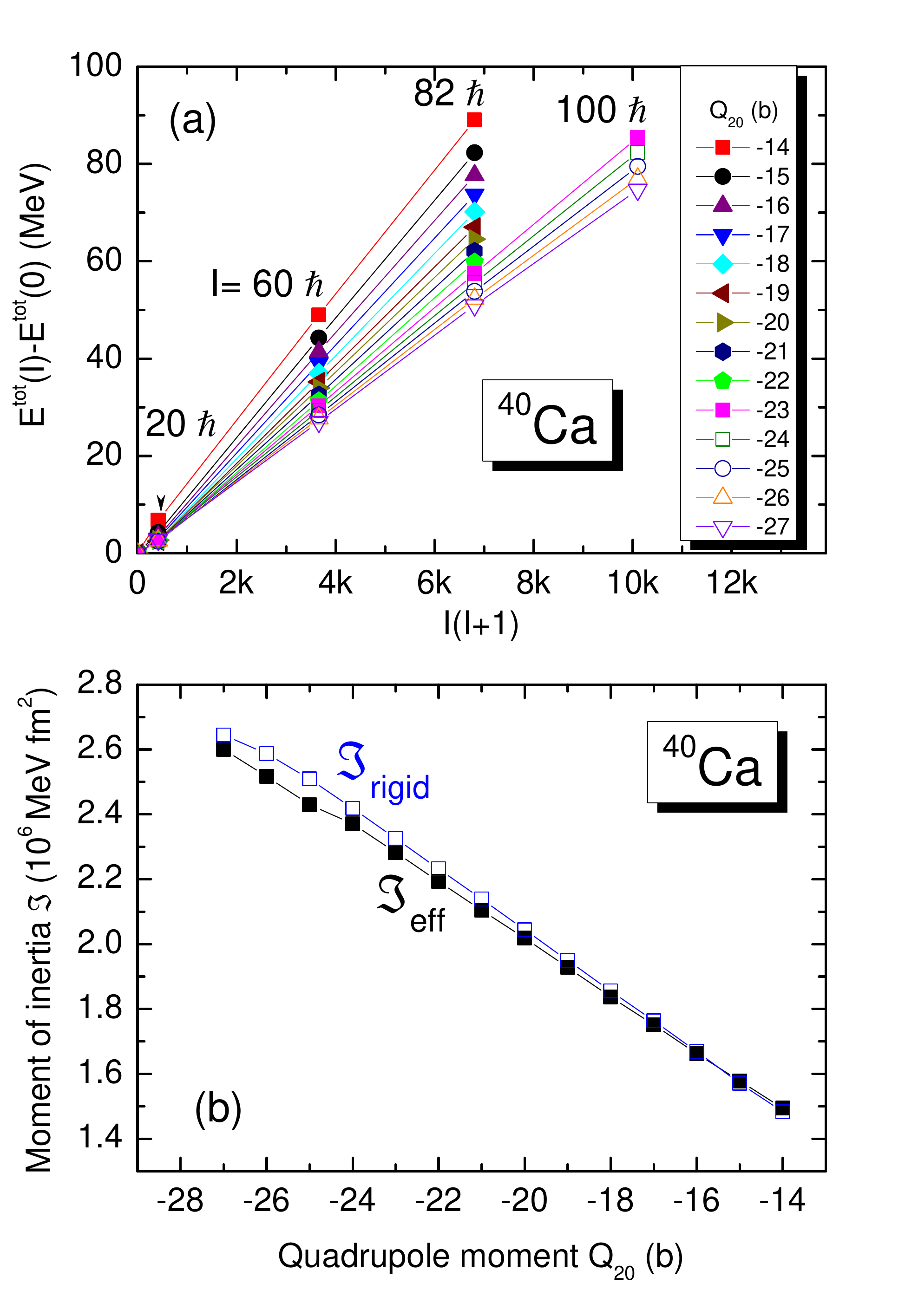}
  \caption{\label{Fig5} (Color online) (a) The total energy difference
    $E^{\rm tot}(I)-E^{\rm tot}(0)$ as a function of $I(I+1)$ for
    toroidal $^{40}$Ca at different $Q_{20}$. The inverses of the
    slopes of different lines give the effective moments of inertia
    ${\Im}_{\rm eff}$.  (b) The effective moments of inertia
    ${\Im}_{\rm eff}$ and the rigid body moments of inertia
    ${\Im}_{\rm rigid}$ as a function of $Q_{20}$ for toroidal
    $^{40}$Ca.}
\end{center}
\end{figure}

To gain new insights into the nature of the non-collective rotational motion,
we determine an effective moment of inertia ${\Im}_{\rm eff}$ for
toroidal $^{40}$Ca from the total energy of the system as a function
of $I$ as $E^{\rm tot}(I)=E^{\rm tot}(0) +{I(I+1)}/{2{\Im}_{\rm eff}}$.
Using the results in Fig.~\ref{Fig3}, we find in
Fig.~\ref{Fig5}(a) that such a linear dependence between
$E^{\rm tot}(I)-E^{\rm tot}(0)$ and $I(I+1)$ holds for different $Q_{20}$.
An effective moment of inertia ${\Im}_{\rm eff}$ can be extracted as a
function of $Q_{20}$.  On the other hand, for different $Q_{20}$, one
can calculate the rigid-body moment of inertia
$\Im_{\rm  rigid}=m_{N}2\pi^{2}R (d^{2}/\ln2) \rho_{\rm max}[R^{2}+3/(2\ln2)d^{2}]$
from the density distributions in Fig.~\ref{Fig4}.  The comparison of
${\Im}_{\rm eff}$ and $\Im_{\rm rigid}$ in Fig.~\ref{Fig5}(b)
indicates the approximate equality of
${\Im}_{\rm eff}$ and $\Im_{\rm rigid}$.
This is in agreement with the result of Bohr and Mottelson who showed
that the moment of inertia associated with the alignment of single-particle
orbits along an axis of symmetry is equal
to the rigid-body moment of inertia \cite{Boh75} and
justifies the use of $\Im_{\rm rigid}$ in the earlier liquid-drop model
of a rotating toroidal nucleus in \cite{Won78}.

It is clear from Fig.~\ref{Fig1}(b) that large shell effects are
expected for some odd $N$ and $Z$ at various $I$ values, and for
combining different $(N,I_{N})$ with $(Z,I_{Z})$ at the same $\hbar
\omega$. Hence light toroidal nuclei with odd-$N$, odd-$Z$, and $N$$\ne$$
Z$ may be possible.  The large shell gaps for $(N, I)$=(58, 58 $\hbar$),
(64, 32 $\hbar$), and (64, 96 $\hbar$) calls for future exploration of
high-spin toroidal isomers in the mass region of $A\sim$120.

In conclusion, under the considerations of the aligned single-particle
angular momentum and the bulk behaviour, the constrained
self-consistent Skyrme-Hartree-Fock model calculations reveal that
high-spin toroidal isomers may have general occurrences in the
mass region of 28$\leqslant$$A$$\leqslant$48.  Experimental search for
these nuclei may allow the extraction of the bulk properties of this
new type of nuclear fluid and its possible utilization as a source of energy.

\section*{Acknowledgments}
The authors wish to thank Drs. Jerzy Dudek, Vince Cianciolo, and I-Yang
Lee for helpful discussions.  This work was supported by the National
Science Center (Poland) and in part by the Division of Nuclear
Physics, U.S. Department of Energy.

\section*{References}


\end{document}